# On the Optimal Design of Triple Modular Redundancy Logic for SRAM-based FPGAs

F. Lima Kastensmidt[1,3], L. Sterpone[2], L. Carro[3], M. Sonza Reorda[2]

[1]Universidade Estadual do Rio Grande do Sul (UERGS)
Engenharia em Sistemas Digitais
Guaíba, Brazil
fernanda-lima@uergs.edu.br

[2]Politecnico di Torino
Dipartimento di Automatica e Informatica
Torino, Italy
luca.sterpone@polito.it,
matteo.sonzareorda@polito.it

[3]Universidade Federal do Rio Grande do Sul (UFRGS)
PPGC – II - DELET
Porto Alegre, Brazil
carro@eletro.ufrgs.br
fglima@inf.ufrgs.br

**Abstract**

*Triple Modular Redundancy (TMR) is a suitable fault tolerant technique for SRAM-based FPGA. However, one of the main challenges in achieving 100% robustness in designs protected by TMR running on programmable platforms is to prevent upsets in the routing from provoking undesirable connections between signals from distinct redundant logic parts, which can generate an error in the output. This paper investigates the optimal design of the TMR logic (e.g., by cleverly inserting voters) to ensure robustness. Four different versions of a TMR digital filter were analyzed by fault injection. Faults were randomly inserted straight into the bitstream of the FPGA. The experimental results presented in this paper demonstrate that the number and placement of voters in the TMR design can directly affect the fault tolerance, ranging from 4.03% to 0.98% the number of upsets in the routing able to cause an error in the TMR circuit.*

## 1. Introduction

*Triple Modular Redundancy* (TMR) is a well-known fault tolerant technique for avoiding errors in integrated circuits. The TMR scheme uses three identical logic blocks performing the same task in tandem with corresponding outputs being compared through majority voters. TMR is especially suitable for protecting designs synthesized in SRAM-based *Field Programmable Gate Arrays* (FPGAs) [1], due to the peculiar effect of an upset in the user's combinational logic design.

In a SRAM-based FPGA, both the user's combinational and sequential logic are implemented in programmable complex logic blocks (CLBs), which are customizable by writing in the SRAM cells of the program memory. When a charged particle strikes one of the sensitive nodes of a cell in such a memory, as a drain in an off state transistor, it generates a transient current pulse that can turn the gate of the opposite transistor on. The effect can produce an inversion in the stored value (in other words, a bit flip in the memory cell). This is called *Single Event Upset* (SEU), or simply upset. The value of an SRAM cell affects either the CLB logic function (by implementing it in a *Lookup Table* or LUT), or the connections among CLBs (by customizing the routing). Changing the value of such a cell has a transient effect followed by a permanent effect on the controlled logic or routing [2]. The upsets in the routing represent the main concern, as 90% of the SRAM cells inside the FPGA control the routing. The main effects of an upset in the routing are open wires and shortcuts between distinct wires. When an upset in the routing affects signals from two distinct redundant logic parts, for example, provoking an undesirable connection between redundant logic part 0 and redundant logic part 1, the TMR approach may not vote the correct output. The probability of this occurrence depends on the logic placement and the number of majority voters in the design. Previous results from bitstream fault injection [3, 4] and radiation ground testing [5] showed that there are few upsets in the routing that can generate an error in the output. This work focuses on how to avoid these types of upsets.

The majority voters perform a very important task in the TMR approach, because they are able to block the effects of an upset through the logic at the final output. In this way, the voters can be placed in the end of the combinational and sequential logic blocks, creating barriers for the upset effects. The problem is to determine the optimal partition of the TMR logic that must be voted inside the circuit in order to reduce the probability that upsets in the routing affect two distinct redundant parts that are voted by the same voters. A small size block partition requires a large number of majority voters that may be too costly in terms of area and performance. On the other hand, placing only voters at the last output increases the probability of an upset in the routing affecting two distinct redundant logic parts overcoming the TMR.

The goal of this paper is to investigate the optimal partition of the TMR logic that must be voted in such a way that the minimal number of voters is used and the maximal upset tolerance is achieved. The considered case study circuit is a digital filter. This design has a large amount of combinational and sequential logic corresponding to dedicated multipliers and adders and registers present in each tap of the filter. To evaluate the TMR robustness a previously developed upset injection tool [6] was used to randomly inject faults in the FPGA bitstream. Results show that there is an optimal solution in terms of number and placement of voters and robustness, which in the case of the filter can reduce from 4.03% to 0.98% (four times smaller!) the number of upsets in the routing able to cause an error in the TMR.

## 2. Related Work

The correct implementation of a TMR circuitry into a SRAM-based FPGA depends on the type of data structure to be protected. As stated in [1], the logic may be grouped into four different structure types: Throughput Logic, State-machine Logic, I/O Logic, and Special Features (Select block RAM, DLLs, etc.). The Throughput Logic is a logic module of any size or functionality, synchronous or asynchronous, where all of the logic paths flow from the inputs to the outputs of the module without ever forming a logic loop. In this case, it is only necessary to triplicate the logic, creating three redundant logic parts (tr0, tr1 and tr2), as presented in fig. 1.

If an upset occurs in one of the redundant combination logic parts (LUTs or routing), its effect will remain until the load of the next bitstream. The constant reconfiguration of the device avoids the accumulation of upsets in the programmable matrix. This continuous loading of the bitstream is called *scrubbing,* and it does interrupt the application. It is important to notice that in Throughput Logic structures composed by registers, the only way to correct an upset in a register is by loading a new data in the input of the register, or by implementing this refreshing structure with voters (see fig. 2). In the case of the registers, it is not possible to load the configuration bitstream without interrupting the application because the correct state of the register cannot be saved and loaded by the bitstream.

State-machine logic is any structure where a registered output, at any register stage within the module, is fed back into any prior stage within the module, forming a registered logic loop. This structure is used in accumulators, counters, or any custom state-machine or state-sequencer where the given state of the internal registers is dependent on its own previous state. In this case, it is necessary to triplicate the logic and have majority voters in the outputs. The register cannot be locked in a wrong value, and for this reason there is a voter for each redundant logic part in the feedback path, making the system able to recover by itself. The same structure presented in fig. 2 can be used. One majority voter can be implemented by one LUT. Because the LUT can be upset (permanent effect), the voters are also triplicated. In this way, if one voter is upset, there are still two voters working properly.

The primary purpose of using a TMR design methodology is to remove all single points of failure from the design. This begins with the FPGA inputs. If a single input was connected to all three redundant logic paths within the FPGA, then a failure at that input would cause these errors to propagate through all the redundancies, and thus the error would not be mitigated. Therefore, each redundant logic part of the design that uses FPGA inputs should have its own set of inputs, as shown in fig. 1. Thus, if one of the inputs suffers a failure, it will only affect one of the redundant logic parts. The outputs are the key to the overall TMR strategy. Since the full triple module redundancy generates every logic path triplicated, the TMR output majority voters, inside the output logic block, allow converging the output again to one signal outside the FPGA, as presented in fig. 1.

There are mainly four types of upsets that may occur in designs synthesized into the FPGA matrix, whose effects are summarized in table 1. They can be classified by the upset location: upsets in the logic (*LUT*), upsets in the customization routing bits inside the CLB (*customization logic in general*), upsets in the routing connecting CLBs and pins (*routing*), and upsets in the CLB flip-flops (*flip-flops*).

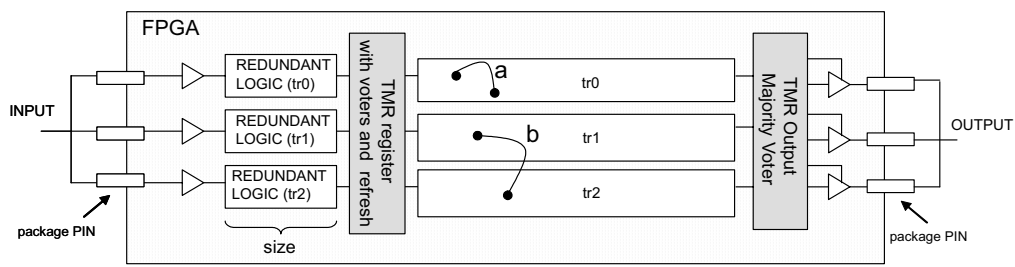

**Figure 1. Triple Modular Redundancy (TMR) Scheme in the FPGA**

The percentages of each type of SRAM cells in the whole set of customizable elements in the programmable matrix of our filter are as follows: the LUTs represent 7.4%, the flip-flops represent 0.46%, the customization bits in the CLB represent 6.36% and the general routing represents 82.9%. Based on these values, one can conclude that the probability of an upset affecting the registers is very low, compared to the probability of this same upset affecting the routing, which is our main concern as the other type of upsets are protected by TMR.

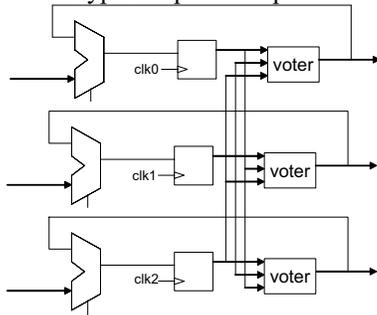

**Figure 2- TMR register with voters and scrubbing**

The probability that an upset in the routing overcomes the TMR is related to the routing density and logic placement. In fig. 1, there are two examples of upsets in the routing. Upset "a" connects two signals from the same redundant part, which does not generate an error in the TMR output, because the upset effect will be voted by the outermost voters. However, upset "b" may provoke an error in the TMR output, because it connects two signals from distinct redundant logic parts affecting two out of three redundant parts of the TMR.

Dedicated floorplanning for each redundant part of the TMR can reduce the probability of upsets in the routing affecting two or more logic modules, but it may not be sufficient, since placement can be too complex in some cases. Remember that each time it is necessary to include voters, there are connections between the redundant parts (see fig. 2), which make impossible to place the redundant logic parts very far away from each other with no connections at all.

If the redundant logic parts tr0, tr1 and tr2 (represented in fig. 1 after the TMR register with voters and refresh) were partitioned in smaller logic blocks with voters, a connection between signals from distinct redundant parts could be voted by different voters. This logic partition by voters is represented in fig. 3. Notice that now the upset "*b*" can not provoke an error in the TMR output, which increases the robustness of the TMR in the presence of routing upsets without being of concern to floorplanning. The problem is to evaluate the best size of the logic to achieve the best robustness. If the logic is partitioned in very small blocks, the number of voters will increase dramatically, causing an overly costly TMR implementation. The objective is finding the best partition in terms of area cost, performance and robustness.

**Table 1. Upset analysis in the Triple Modular Redundancy approach**

| Upset Location | Upset Effect | Consequences | Upset Correction |
| --- | --- | --- | --- |
| LUT | Modification in the Combinational logic | - Error in one redundant part with no error in the TMR design output | By scrubbing |
| Routing | Connection or disconnection between any two or more signals in the design | - Error in the redundant part with no error in the TMR design output<br>- Error in more than one redundant part with error in the design output | By scrubbing |
| Customization logic in general | Connection or disconnection between any two signals in the same CLB | - Error in the redundant part with no error in the TMR design output<br>- Error in more than one redundant part with error in the design output | By scrubbing |
| Flip-flops | Modification in the sequential logic | - Error in the redundant part, no error in the TMR design output | By design modification |

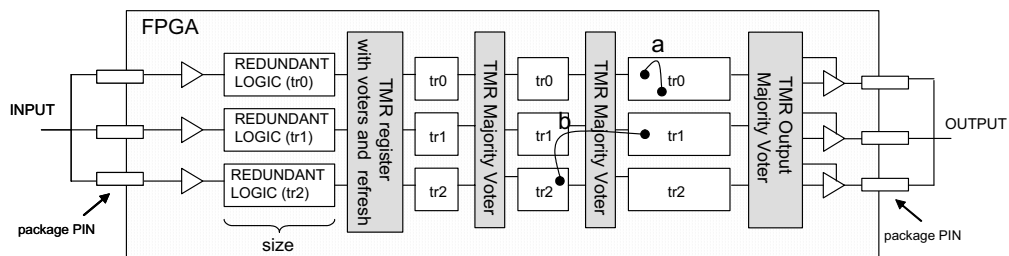

**Figure 3. Triple Modular Redundancy (TMR) scheme with logic partition in the FPGA**

## 3. The Case Study Design: Digital Fir Filter

The designed test case to evaluate the robustness of the TMR according to the logic partition is an 11 taps 9-bit digital low-pass filter. The original coefficients calculated by Matlab [7] were multiplied by the constant 512. The final multiplier coefficients are: 1, -1, -9, 6, 73 and 120. In this filter design, there are eleven dedicated 9-bit multipliers, ten 18-bit adders and ten 9-bit registers.

An upset can affect the registers (in this case causing a transient effect), or can affect the logic (multipliers, adders, voters), causing a permanent effect. Five different versions of the filter were implemented. The first is the standard one with no protection at all (*filter*). The second version was protected by TMR using the maximum logic partition (*TMR_p1*). In this case, each combinational logic component, such as an adder or a multiplier, was triplicated and majority voters were inserted in the output as presented in fig. 4(a). The third implementation is a TMR filter, where each partition logic block has one adder, one multiplier and voters placed in the output (*TMR_p2*), as shown in fig. 4(b). The fourth TMR filter version has only majority voters in the outermost output signals (*TMR_p3*), as presented in fig. 4(c).

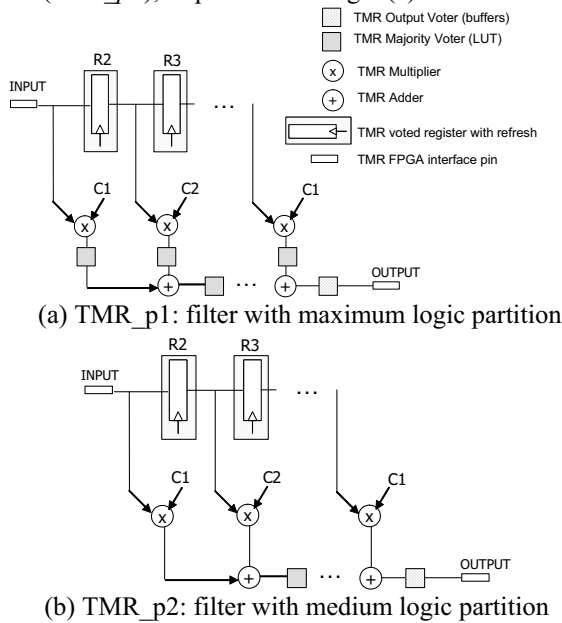

(a) TMR_p1: filter with maximum logic partition

(b) TMR_p2: filter with medium logic partition

(c) TMR_p3: filter with minimum partition

**Figure 4. TMR digital filter schemes used to evaluate the optimal logic partition of TMR**

All the TMR implementations so far have TMR registers with voters and refresh. The last implementation differs from the fourth version because the registers in this case are not voted (*TMR_p3_nv*). They are just triplicated and the final majority voters are responsible for voting the upsets in the combinational logic (adders and multipliers) and in the sequential logic (registers), as well as the routing upsets.

The comparison between the main characteristics of the four implemented versions of the FIR digital filter are reported in Table 2, where *slice* reports the number of FPGA slices that the circuit uses, *bitstream* is composed of the *routing bits*, *LUTs bits* and *CLB flip-flops*. The *routing bits* is the number of configuration memory bits for interconnections and multiplexers used for routing the signals of the circuits, *LUTs bits* is the number of configuration memory bits used to program the LUTs within the CLBs, and *CLB flip-flops* is the number of configuration memory bits used to program the flip-flops within the CLBs.

The amount of *routing bits* in the unprotected design corresponds roughly to 80% of the total customizable bits. In the TMR version, the design with the maximum partition has 77% of the total customizable bits corresponding to *routing bits*, whereas the minimum partition presents 81% of *routing bits*. This result is very interesting, because it shows that increasing the logic partition with voters actually decreases the percentage of *routing bits* compared to the logic. This can happen because of the amount to logic that is inserted by the voters.

**Table 2. Comparison between TMR partitioned designs in XC2S200E-PQ208**

| Filter Design | Area (# slices) | Bitstream | | | Estimated Performance |
|---|---|---|---|---|---|
| | | #routing bits | #LUTs bits | #CLB ffps bits | |
| Standard: no TMR protected | 150 | 42,953 | 9,600 | 722 | 154 Mhz |
| TMR: max. partition (TMR_p1) | 560 | 138,453 | 35,840 | 3,498 | 123 Mhz |
| TMR: med. partition (TMR_p2) | 504 | 161,568 | 32,256 | 3,492 | 137 Mhz |
| TMR: min. partition (TMR_p3) | 498 | 151,994 | 31,872 | 3,447 | 153 Mhz |
| TMR: min. partition / no voted registers (TMR_p3_nv) | 476 | 150,521 | 30,464 | 2,141 | 154 Mhz |

## 4. Fault Injection Experiments

The robustness of the TMR technique implemented in a high-level description language synthesized in the Spartan® FPGA was evaluated by injecting faults in the configuration bits of the matrix (LUTs and configuration routing cells). We used a fault-injection system we developed [6] which is composed of three modules:
1. *Fault List Manager*: it generates the list of faulty bitstreams to be injected within the circuit under analysis.
2. *Fault Injection Manager*: it manages the fault injection process, by selecting one fault from the fault list, performing its injection in the circuit under analysis, and then observing and analyzing the obtained results to provide the fault classification.
3. *Golden device*: it is a copy of the DUT without the usage of TMR.

The fault injection system consists of the Fault List Manager implemented as a software process that runs on a PC, a Fault Injection Manager running in part on a PC as a software process, and in part on the same FPGA where the circuit under analysis is placed, and finally a Golden device placed on the same FPGA, and linked with the Fault List Manager and with the DUT. Every clock cycle their output signals are compared with those coming from the DUT.

In order to improve the fault-injection process, we designed the *Fault List Manager* so that it is able to identify the configuration memory bits that are actually programmed to implement the DUT and generate the bit-flips only for them. To implement this solution we first generated the file storing the configuration memory file for the whole design; secondly the Fault List Manager identified those bits related to the DUT. Please note that faults were selected in such a way that common-mode errors were not possible. This process is possible thanks to a data base of the programmed resources (LUTs and configuration routing cells) we developed by decoding the Xilinx bitstream. The list of faults for the DUT is finally computed and stored. Each element of the fault list is a faulty bitstream for the FPGA, where one bit at a time is modified emulating the effect of a SEU.

In the used fault-injection system the DUT and the Fault Injection Manager implemented on the FPGA share the same FPGA device. However, to guarantee that the modules are placed on the FPGA device in such a way that any fault injected in the DUT does not interfere with the Fault Injection Manager, we constrained the place and route tools to organize the FPGA-resource floorplanning allocation. The DUT and the Fault Injection Manager are placed separately on the floorplanning by at least 10 slice columns to avoid the link between the two units generated by an undesirable interconnection due to an upset.

The FPGA device used in our experiment is a Spartan® XC2S200EPQ208 from Xilinx [8], whose configuration memory is composed of 1,442,016 bits organized in 2,501 frames of 576 bits each. The configuration memory controls an array of 28 x 42 slices. The robustness of the TMR technique implemented in a high-level description language was observed while running five fault-injection campaigns, one for each circuit.

The fault injection process took about 13 seconds for each fault injected. The test pattern generation and the output analyzer placed in the same FPGA need a negligible time for applying all the input stimuli and classifying the faults (5.7 ms on the average). Most of the time for evaluating each injected fault was due to the time for downloading the faulty bitstream within the SRAM-based FPGA. The results we gained performing the fault-injection campaigns are reported in table 3, where *Injected Faults* reports the number of SEUs injected, and *Wrong Answer* reports the percentage of SEUs provoking a fault within the TMR DUT.

Roughly 10% of the whole configuration memory bits were injected. This set was randomly chosen from the fault list. It would take approximately 3 months for testing the whole configuration memory bits of the five circuits. The results confirm our original hypothesis, since they show that placing voters just at the final output is not sufficient to avoid errors. There is an optimum size of triplicated logic that must be voted inside the design, creating voter barriers along the logic that help block the effect of upsets in the routing to propagate to the output. The results experimentally demonstrate that the logic partition done in the TMR_p2 version reduce the percentage of uncovered faults from 4.03% to 0.98%, corresponding to a four times reduction. This solution is also efficient in term of estimated performance, as shown in Table 2.

**Table 3. Fault injection campaign results**

| Design | Injected Faults [#] | Wrong answer [#] | [%] |
|---|---|---|---|
| Standard Filter | 5,100 | 4,952 | 97.10 |
| TMR_p1 | 17,515 | 706 | 4.03 |
| TMR_p2 | 19,401 | 190 | 0.98 |
| TMR_p3 | 18,501 | 289 | 1.56 |
| TMR_p3_nv | 18,000 | 2,268 | 12.60 |

Table 4 reports a classification of the effects of the injected upsets that caused an error in the TMR. The effect analysis of each faulty bitstream was done using the classification tool developed in [9]. The results confirm that the routing resources are the most sensitive to upsets. No upsets in the LUTs could provoke an error in the TMR design, as it was expected. Upsets in the CLB multiplexers (mux) and in the initialization that are related to the routing resources represent less than 8% of the total effects. The main problem is the general routing that connects CLBs and I/O pads, corresponding to more than 90% of routing resources. The general routing effects that can provoke an error in the TMR output are classified as *Open, Bridge, Input-Antenna, Conflict* and others. The *Open* effect corresponds to a Programmable Interconnection Point (PIP) in open state (open connection). The *Bridge* effect means that a new PIP is permitted. Both Bridge and Open effects may influence the behavior of the circuit. The *Input-Antenna* effect also represents a new permitted PIP. In this case, this new PIP connects an unused input node to a used output node; this may influence the behavior of the circuit, since the CLB or output pad are driven to an unknown logic value. The *Conflict* effect is when a new PIP links both used input and output nodes, creating a conflict and a propagation of an unknown value along the TMR. These effects are a real challenge for those designers who are involved in devising solutions for hardening their FPGA-based circuits, and these effects must be avoided or at least minimized.

It is important to notice that one upset can provoke more that one effect in the programmable matrix. The same can be said about different upsets in the bitstream that can present the same effect. This explains why the number of table 3 and 4 are not exactly the same.

## 5. Conclusions

The results presented in this paper suggest that there is a trade off between the logic partition of the throughput logic (and consequently between the number of voters) and the number of routing upsets that could provoke an error in the TMR. In contrary to what was expected, a large number of voters does not always mean larger protection against upsets. There is an optimal logic partition for each circuit that can reduce the propagation of the upset effect in the routing. For the case study circuit, the best partition is the medium partition (TMR_p2). This version of the TMR design presents a small sensitivity to routing upsets (0.98%, a four times improvement over normal TMR) and small performance degradation (about 10%) compared to the standard version (not protected).

Future works will consider analyzing the effect of logic partition in power dissipation, and resolving the problem of the uncovered routing upsets by means of a combination of logic partition and dedicated floorplanning.

**Table 4. Effects induced by the injected upsets**

| | | Standard | | TMR_p1 | | TMR_p2 | | TMR_p3 | | TMR_p3_nv | |
|---|---|---|---|---|---|---|---|---|---|---|---|
| | | [#] | [%] | [#] | [%] | [#] | [%] | [#] | [%] | [#] | [%] |
| CLBs Logic and Routing | LUT | 852 | 16 | 0 | 0 | 0 | 0 | 0 | 0 | 0 | 16 |
| | MUX | 123 | 2 | 16 | 2 | 1 | 0 | 15 | 4 | 367 | 2 |
| | Initialization | 174 | 3 | 13 | 1 | 0 | 0 | 11 | 3 | 400 | 3 |
| General Routing | Open | 1321 | 25 | 276 | 38 | 82 | 40 | 126 | 37 | 1672 | 25 |
| | Bridge | 427 | 8 | 62 | 9 | 41 | 20 | 42 | 12 | 403 | 8 |
| | Input-Antenna | 76 | 1 | 33 | 5 | 7 | 3 | 14 | 4 | 73 | 1 |
| | Conflict | 1342 | 25 | 26 | 4 | 13 | 6 | 6 | 2 | 185 | 25 |
| | Others | 1006 | 20 | 301 | 41 | 66 | 31 | 128 | 38 | 756 | 20 |
| | Total | 5321 | | 727 | | 210 | | 342 | | 3856 | |